\documentclass[aps,prb,twocolumn,superscriptaddress,groupedaddress,longbibliography]{revtex4-1}
\usepackage{epsfig,amsopn}
\usepackage{graphicx}
\usepackage{color}
\usepackage{amsmath,amssymb}
\usepackage{enumerate}
\newcommand\bea{\begin{eqnarray}}
\newcommand\eea{\end{eqnarray}}
\newcommand\beq{\begin{equation}}
\newcommand\eeq{\end{equation}}

\def\nn{\nonumber}
\def\f{\frac}

\def\De{\Delta}
\def\dg{\dagger}

\begin{document}
\title{Probing nonlocality of Majorana fermions in Josephson junctions of Kitaev chains connected to normal metal leads} 
\author{ Abhiram Soori~~}
\email{abhirams@uohyd.ac.in}
\affiliation{ School of Physics, University of Hyderabad, C. R. Rao Road, Gachibowli, Hyderabad-500046, India.}

\begin{abstract} 
Kitaev chain~(KC) is a prototypical model for the study of Majorana fermions~(MFs). In the topological phase,
a KC hosts two MFs at its ends. Being separated in space, these two MFs are nonlocal. 
The nonlocal transport in a KC biased between two normal metal leads is mediated by 
electron tunneling~(ET) and crossed Andreev reflection~(CAR). ET contributes 
positively while CAR contributes negatively to the  nonlocal conductance. Enhanced CAR and hence a negative  
nonlocal conductance is a hallmark of nonlocality of MFs. But simple conductance measurements in 
the above setup cannot probe the nonlocality of MFs due to the almost cancellation of currents from ET and CAR. 
On the other hand, a Josephson junction between two KCs hosts two Andreev bound states~(ABSs) at 
the junction formed by a recombination of Majorana fermions of the individual KCs. The energies of 
the ABSs are away from zero and can be changed by altering the superconducting phase difference. A Josephson 
junction between two finitely long KCs hosts two MFs at the two ends and two ABSs at the junction. 
We show that when normal metal leads are connected to two ends of such a Josephson junction, the nonlocal 
conductance of the setup can be negative for bias values equal to the energies of the ABSs. Thus the 
nonlocal conductance in such setup can probe the nonlocality of the constituent MFs.
\end{abstract}
\maketitle

\section{Introduction}

In 2001, Kitaev showed that in a one-dimensional lattice model with p-wave superconductivity topologically nontrivial
isolated Majorana fermions~(MFs) can exist~\cite{kitaev2001unpaired}. This lattice model known as Kitaev chain~(KC)
attracted huge interest because MFs could be used as building blocks of a topological quantum computer~\cite{nayak08}.
Interestingly, four decades prior to the work of Kitaev, it was shown by Lieb, Shultz and Mattis that anisotropic
Heisenberg spin chain can be solved by mapping it to p-wave superconducting chain of spinless fermions~\cite{lieb1961}. 
In the last decade, an advance in modeling put forward the idea that isolated MFs can be realized in spin orbit coupled 
quantum wires placed in proximity to a s-wave superconductor accompanied by a magnetic 
field~\cite{lutchyn2010majorana,oreg2010helical}. Such MFs should exhibit zero bias conductance peak~(ZBCP)
when a normal metal~(NM) lead is connected. In the years that followed, several experiments convincingly
detected ZBCP upholding the realization of isolated MFs~\cite{mourik2012,Das2012,albrecht2016,Zhang2018,aguado17}. 
However, there are two MFs in a KC and they are spatially separated. This nonlocal aspect of MFs though already 
noticed by Kitaev himself~\cite{kitaev2001unpaired} has
not been studied experimentally. 
In the limit of infinite length of the KC, the two MFs are decoupled
and are exactly at zero energy, but for a finite length of the KC the two MF wavefunctions overlap
leading to the formation of two nonlocal Dirac fermions (this happens generically except for a special choice of 
parameters). The nonlocal transport in NM-KC-NM
is mediated by electron tunneling~(ET) and crossed Andreev reflection~(CAR). In the former~(latter),
an electron incident from one NM tunnels through the KC and exits onto another NM
as an electron~(a hole). The local transport is mediated by electron reflection~(ER) and Andreev reflection~(AR). 
In the former~(latter), the electron incident from one NM results in a reflected electron~(hole) in
the same NM. Local conductance is the differential conductance at the first NM-KC junction.
Nonlocal conductance or transconductance is the differential conductance between the first NM and the 
second NM maintaining the KC and the second NM grounded. AR is a definite signature
of isolated MFs and a local conductance of $2e^2/h$ at zero bias
owing to perfect AR is obtained when the two MFs in the KC are decoupled. The isolated
MF at one end of the KC is responsible for perfect AR. CAR on the other hand is a nonlocal process
mediated by both the MFs in the chain that fuse to result in a nonlocal Dirac fermion. Though ET
is also mediated by the two MFs, ET does not need spatially separated MFs. An enhanced CAR over ET is 
a hallmark of nonlocality of the MFs. However, the currents due to CAR and ET almost cancel out in a 
simple setup made with two NMs connected to a KC~\cite{liu17,nilsson2008}. This has 
motivated proposals to probe nonlocality of MFs by noise measurements~\cite{nilsson2008,Lu12,Liu13,Zocher13,Prada17}. 
Also, we recently proposed to a setup consisting of a Kitaev ladder~\cite{nehra19} in series with Kitaev 
chain which can probe the nonlocality of MFs by measurement of nonlocal conductance~\cite{soori19trans}. 
In this work, we showed that the negative nonlocal conductance  at energies of the nonlocal Dirac 
fermion is the hallmark of the nonlocality of MFs. In addition, there are many theoretical works that 
study the nonlocal aspect of Majorana fermions~\cite{fu10,zazunov11,hutzen12,zazunov12,wang13,sau15,beri12,
altland13,beri13,altland14,vanBeek16,bao17,vanBeek18,law09}.
Another setup that can enhance CAR over ET will not
only probe the nonlocality of MFs but  will also add to an assortment of many existing proposals to enhance
CAR over ET~\cite{deutscher2000,he14,yeyati07,soori17,nehra19,soori19trans,liu14,Chen2015}. In this respect,
an interesting proposal to enhance CAR is by Chen~et.~al. where the phase in the superconducting channel is 
varied continuously resulting in a negative nonlocal conductance~\cite{Chen2015}. This is essentially a 
series of Josephson junctions connected to two NM leads. Motivated by this proposal in s-wave superconductors, 
we expect that in a single Josephson junction made by p-wave superconductors connected to NM leads, CAR can be 
enhanced over ET. 
 
\begin{figure}
  \includegraphics[width=9cm]{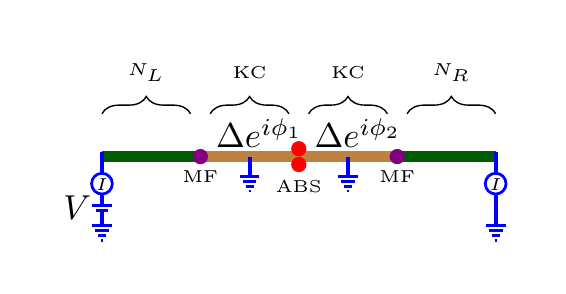}
  \caption{Schematic of the proposed setup. Two KCs form a Josephson junction at the center.  
  Two NMs  one on left~($N_L$) and one on right~($N_R$) connects the JJ. A bias $V$ is applied to $N_L$. 
  The Josephson junction and $N_R$ are grounded. Current meters $I$ are inserted in $N_L$ and $N_R$ to measure the currents 
  $I_L$ and $I_R$ respectively.}~\label{fig-schem}
 \end{figure}

 In this paper, we propose to connect NM leads to a Josephson junction of two KCs as shown in
 Fig.~\ref{fig-schem}. Such a Josephson junction has already been realized experimentally~\cite{rokhi2012} and 
 connecting NM leads to such a Josephson junction appears to be an easy task. Josephson junction 
 made out of p-wave superconductors exhibits fractional Josephson effect~\cite{kwon04}. However, the $4\pi$ 
 periodicity in current phase relation does not survive when two topological superconductors of finite length are 
 coupled due to the hybridization of the all four Majorana fermions~\cite{cayao17}. Still the one can say whether 
 the topological superconductor is in the topologically trivial phase or non-trivial phase by studying the dependence 
 of the critical current of the Josephson junction on the length of the superconductor or on the junction 
 transparency~\cite{cayao17}.  A superconducting phase difference 
 drives a Josephson current between the two superconductors, but we are interested in the currents that flow in 
 the attached NMs due to an applied bias. The paper is structured as follows. In the next section we 
 discuss the details of calculation, which is followed by the section that discusses the results. Finally we end 
 the paper with a concluding section. 
 
 \section{Calculations}
 The Hamiltonian for the proposed system is
 \bea H&=&H_L+H_{JJ}+H_R+H_{LJ}+H_{JR}, \label{eq:ham}\eea
 where $H_L$ describes the NM on left $N_L$, $H_{JJ}$ describes the Josephson junction made out of the KCs, 
 $H_R$ describes the NM on right $N_R$, $H_{LJ}$ describes the coupling between $N_L$ and the Josephson junction
 and $H_{JR}$ describes the coupling between the Josephson junction and $N_R$. Various terms in the Hamiltonian are 
 \bea 
 H_L &=& -t\sum_{n=-\infty}^{-1}(c^{\dg}_{n-1}c_n+c^{\dg}_{n}c_{n-1}) \nn \\ 
 H_{JJ} &=& \sum_{n=0}^{L-2}[-t(c^{\dg}_{n}c_{n+1}+{\rm h.c.})+\De(e^{i\phi_1}c^{\dg}_{n}c^{\dg}_{n+1} + {\rm h.c.})]\nn\\
 &&-t_{MK}(c^{\dg}_{L-1}c_L+{\rm h.c.}) \nn\\
 &&+\sum_{n=L}^{2L-2}[-t(c^{\dg}_{n}c_{n+1}+{\rm h.c.})+\De(e^{i\phi_2}c^{\dg}_{n}c^{\dg}_{n+1}+{\rm h.c.})]\nn \\
 H_R &=& -t\sum_{n=2L}^{\infty}(c^{\dg}_{n+1}c_n+c^{\dg}_{n}c_{n+1}) \nn \\ 
 H_{LJ}&=& -t_L(c^{\dg}_{-1}c_0+{\rm h.c.}) \nn \\
 H_{JR}&=& -t_R(c^{\dg}_{2L-1}c_{2L}+{\rm h.c.}). \label{eq:ham2}
 \eea
 The Josephson junction is formed by two KCs having superconducting phases $\phi_1$ and $\phi_2$. 
Each KC here consists of $L$ sites. They host  isolated MFs in either of the two limits: $\De=\pm t$ or $L\to\infty$.
In a finitely long Kitaev chain in the topological phase when $t\neq\De$, the two MFs at the ends of a KC hybridize 
 with each other forming two nonlocal Dirac fermions at energies $\pm E_g$, where $E_g$ falls off exponentially
 with the length of the KC. 
 Further, the Josephson junction between the two KCs does not 
 have a superconducting hopping. This means Josephson junction in our model can be imagined to be made of
 a thin insulator region between the two KCs. The two MFs at the junction hybridize and form Andreev bound 
 states~(ABSs) as shown in Fig.~\ref{fig-schem}.
  \begin{figure}
 \includegraphics[width=8cm]{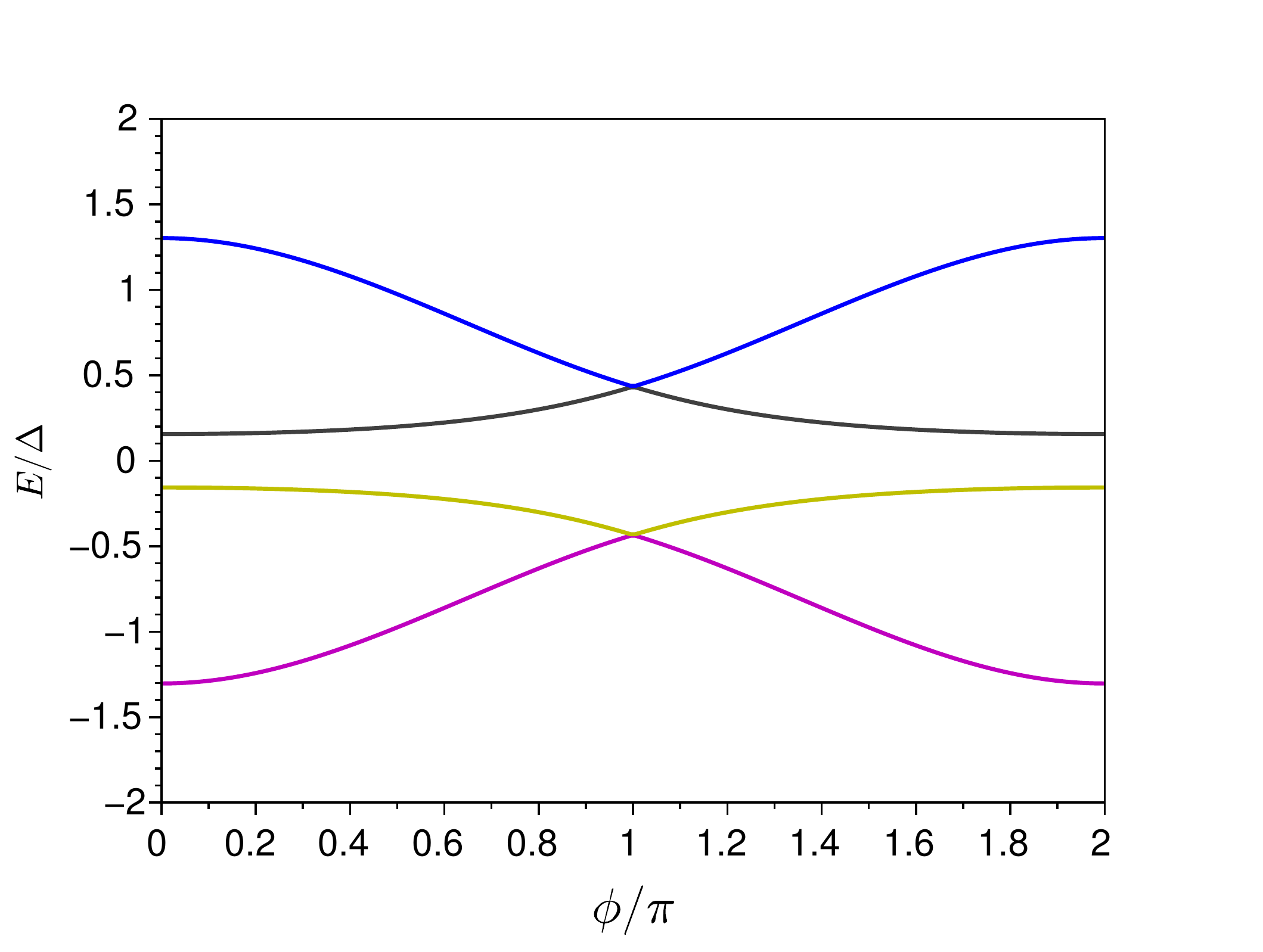}
 \caption{Energy levels of the Josephson junction near zero energy as a function of the phase difference
 $\phi=(\phi_1-\phi_2)$  for the choice of parameters: $\De=0.2t$, $t_{MK}=0.4t$ and $L=10$.}\label{fig01}
\end{figure}
In Fig.~\ref{fig01}, energy levels of the isolated Josephson junction made of two KCs described by the Hamiltonian 
$H_{JJ}$ are plotted as a function of the superconducting phase difference $\phi=(\phi_1-\phi_2)$ for the choice of parameters
$\De=0.9t$ and $t_{MK}=0.4t$. The energy levels closer to zero energy are those of the fermions formed by the hybridization 
of the MFs that live at the ends (away from the junction). The energy levels further away from zero energy but in the range
$(-1.5\De,1.5\De)$ are ABSs that live at the junction as depicted in Fig.~\ref{fig-schem}. These ABSs are formed by the 
hybridization of the MFs at the junction. The energy levels outside $\pm2\De$ are those of the quasiparticles that belong
to the bulk of the KCs and are not shown in the figure. Hence, a conductance spectroscopy in the energy range $(-2\De,2\De)$ 
will probe the constituent MFs of the KC that form the Josephson junction.

For an electron incident from $N_L$ at energy $E$, the wavefunction
$[\psi^e_n,~\psi^h_n]^T$ has the form
\bea 
\psi^e_n&=&e^{ik_ean}+r_ee^{-ik_ean} ~~{\rm for~~}n\le-1\nn \\
&=&t_ee^{ik_ean}~~{\rm for~~}n\ge 2L \nn \\
\psi^h_n&=&r_he^{ik_han} ~~{\rm for~~}n\le-1\nn \\
&=&t_he^{-ik_han}~~{\rm for~~}n\ge 2L,
\eea
where $k_ea=\cos^{-1}[-(E+\mu)/2t]$, $k_ha=\cos^{-1}[(E-\mu)/2t]$. 
Here, $n$ is the site~index and $a$ is the lattice spacing. The scattering coefficients 
$r_e$, $t_e$, $r_h$ and $t_h$ can be determined by solving the equation $H\Psi=E\Psi$ using the full Hamiltonian (eq.~\eqref{eq:ham})
where $\Psi$ is the full wavefunction. An electron incident from $N_L$ at energy $E$ contributes to the local differential conductance
$G_{LL}$ and the differential transconductance $G_{RL}$ at a voltage bias $V=E/e$ (where $e$ is the electron charge). $G_{LL}$
($G_{RL}$) is the ratio of the change in current $dI_L$ ($dI_R$) in $N_L$ ($N_R$) to the change in bias in $N_L$ when the bias
is changed from $V$ to $V+dV$. The two conductances are given by the Landauer-Buttiker formula~\cite{nehra19,soori19trans}
\bea 
G_{LL}&=&\f{e^2}{h}\Big[1-|r_e|^2+|r_h|^2\f{\sin{k_ha}}{\sin{k_ea}}\Big] \nn \\
G_{RL}&=&\f{e^2}{h}\Big[|t_e|^2-|t_h|^2\f{\sin{k_ha}}{\sin{k_ea}}\Big] \label{eq:conductance}
\eea

\section{Results and Analysis}
\begin{figure}
 \includegraphics[width=4cm]{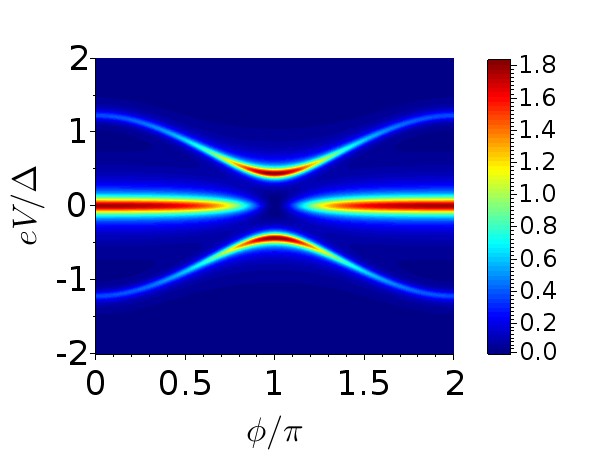}
 \includegraphics[width=4cm]{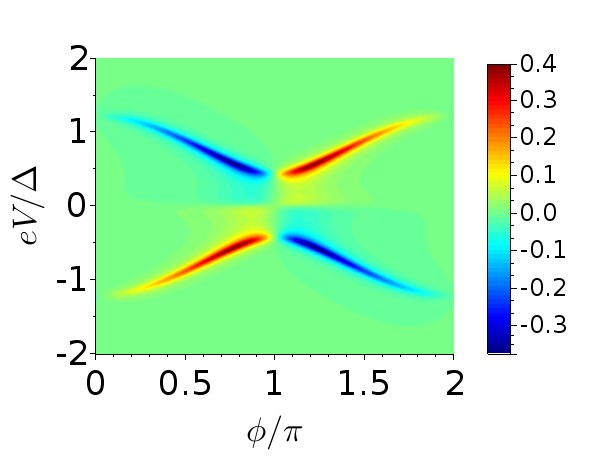}
 \caption{$G_{LL}$ (left panel) and $G_{RL}$ (right panel) in units of $e^2/h$ versus bias $eV$ and the phase difference 
 $\phi=(\phi_1-\phi_2)$  for the choice of parameters: $L=10$, $\De=0.2t$, $t_{MK}=0.4t$, $t_{L}=0.2t$ and $t_{R}=0.9t$.}\label{fig02}
\end{figure}

\begin{figure}
 \includegraphics[width=8cm]{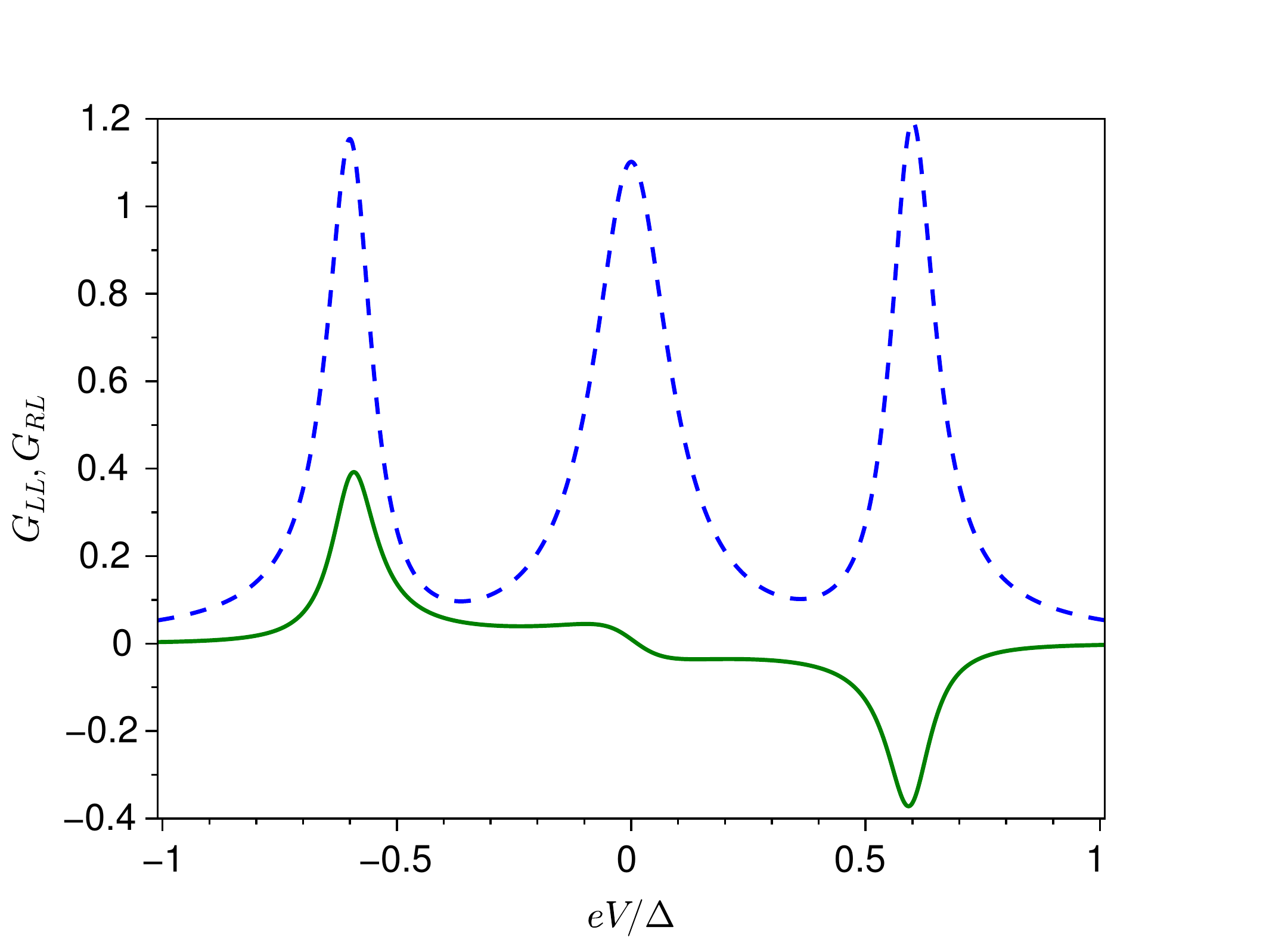}
 \caption{$G_{LL}$ (blue dashed line) and $G_{RL}$ (green solid line) in units of $e^2/h$ versus the bias $eV$ for the same
 choice of parameters as  in Fig.~\ref{fig02}, for $\phi=0.75\pi$. }~\label{fig02sub1}
\end{figure}

\begin{figure}
 \includegraphics[width=8cm]{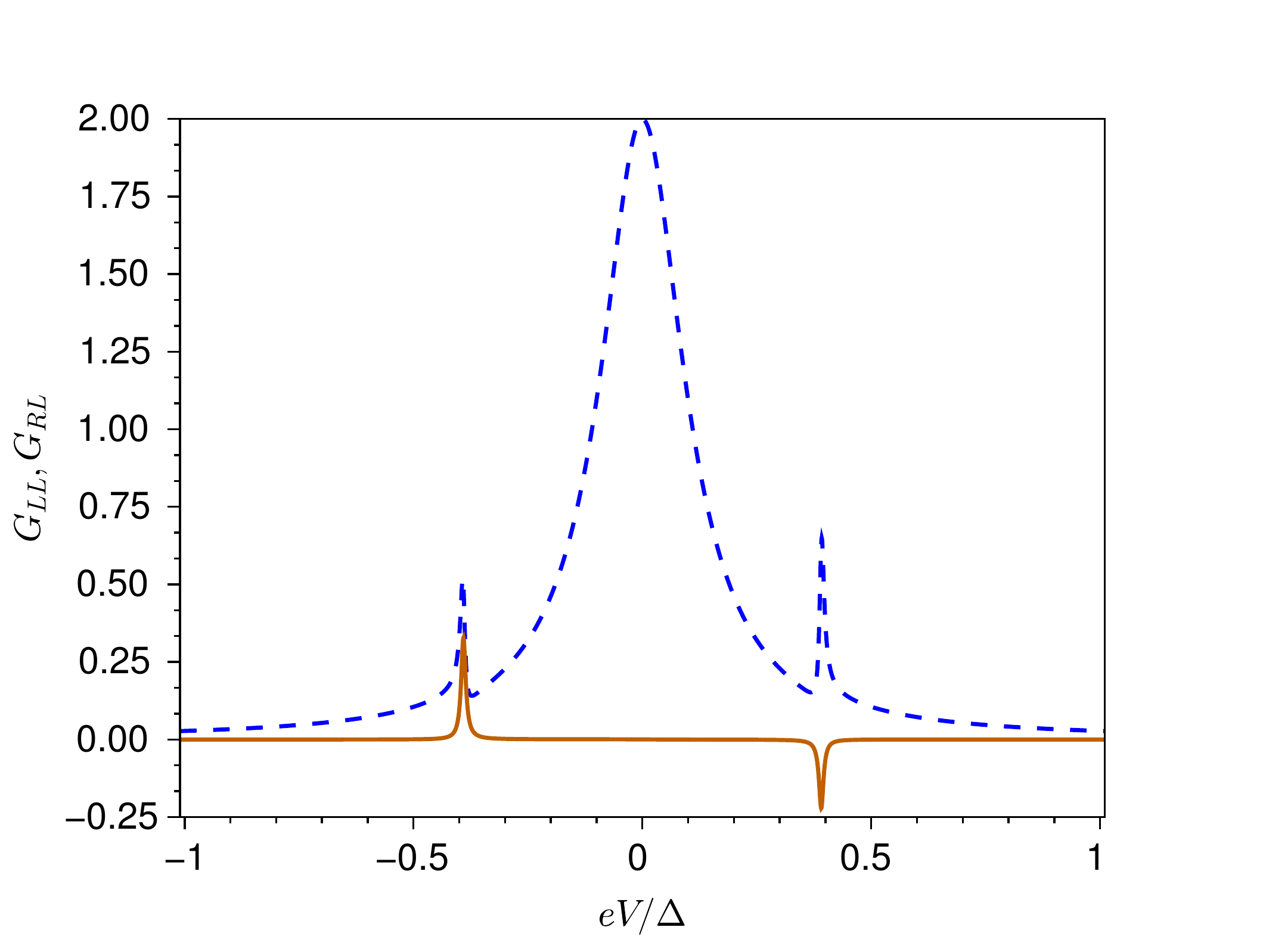}
 \caption{$G_{LL}$ (blue dashed line) and $G_{RL}$ (brown solid line) in units of $e^2/h$ versus the bias $eV$ for the 
 same choice of parameters as  in Fig.~\ref{fig02sub1} except for $L=15$. }~\label{fig02sub2}
\end{figure}

\begin{figure}
 \includegraphics[width=4cm]{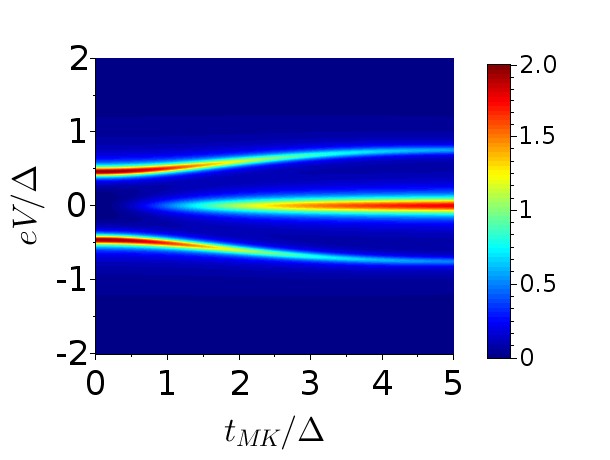}
 \includegraphics[width=4cm]{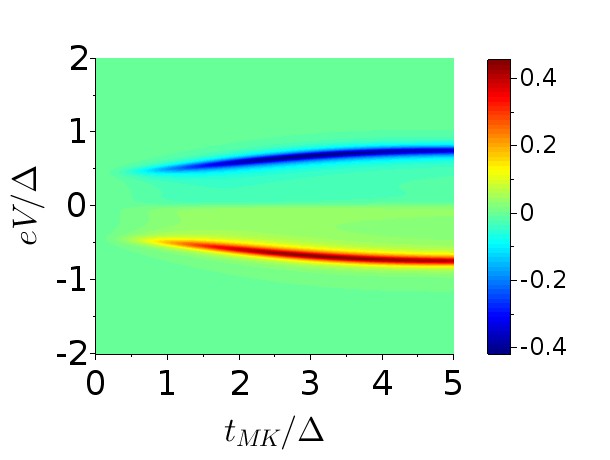}
 \caption{$G_{LL}$ (left panel) and $G_{RL}$ (right panel) in units of $e^2/h$ versus bias $eV$ and the Josephson coupling $t_{MK}$
   for the choice of parameters: $\De=0.2t$, $\phi_1-\phi_2=0.75\pi$,  $t_{L}=0.2t$ and $t_{R}=0.9t$.}\label{fig03}
 \end{figure}

 We calculate the local conductance $G_{LL}$ and the transconductance $G_{RL}$ as a function of the bias $eV$ and the superconducting
 phase difference $\phi=\phi_1-\phi_2$ in Fig.~\ref{fig02} for the choice of parameters $L=10$, $\De=0.2t$, $t_{MK}=0.4t$, $t_{L}=0.2t$ 
 and $t_{R}=0.9t$. We see that the local conductance shows peaks close to zero bias. From Fig.~\ref{fig01}, we can see that these are 
 energy levels of the nonlocal Dirac fermions formed by the hybridization of the MFs that live at the ends, away from Josephson junction. 
 An electron that enters the Josephson junction near zero bias sees the MF at the end and gets Andreev reflected almost perfectly.
 At the energies of the ABSs, we see peak in local conductance but the height of the peak is much smaller~($\sim0.45e^2/h$). However, 
 at the energies of the ABSs, the nonlocal conductance dominates at finite phase difference. The transconductance $G_{RL}$ shows peaks
 and valleys at the energies of ABSs. This is due to the interference effect between the electron excitations and hole excitations at the
 Josephson junction. We see that the transconductance touches negative values high in magnitude ($\sim-0.4e^2/h$) marking an enhanced
 CAR over ET. This is a hallmark of nonlocality of the constituent MFs. The enhanced nonlocal transport at energies of the ABSs for 
 the choice of $\phi$ closer to $\pi$ is because the ABSs formed near the Josephson junction promote nonlocal transport as they 
 have sufficient overlap with the incident electron mode from $N_L$ for this choice of parameters. 
To see the date in the form of line plot, we plot the local conductance and the transconductance as a function of the bias for the
same parameters, fixing $\phi=0.75\pi$ in Fig.~\ref{fig02sub1}. As the length of the each KC is increased, the MFs get more localized and 
the hybridization between the MFs at the extreme ends becomes less prominent. This promotes local transport over nonlocal transport though
the ABSs at the junction are formed by the hybridization of the two MFs close to the junction. This is because the ABSs formed at the 
junction have a much lower overlap with the incident electron from $N_L$. We can  see this in Fig.~\ref{fig02sub2} plotted for the choice\
of same parameters except for a longer length of the KCs, $L=15$. 
 Now, we study the dependence of the $G_{LL}$ and $G_{RL}$ on the bias and the Josephson coupling $t_{MK}$ at a fixed phase difference. In 
 Fig.~\ref{fig03}, we plot $G_{LL}$ and $G_{RL}$ as functions of the applied bias and the Josephson coupling for the choice of parameters 
 same as in Fig.~\ref{fig02}, fixing $\phi=\phi_1-\phi_2=0.75\pi$. At $t_{MK}=0$, the two KCs are decoupled and the MFs in the left
 KC hybridize to form Dirac fermions at energies $\pm0.5\De$ which exhibit perfect Andreev reflection at these energies as can 
 be seen from the left panel of Fig.~\ref{fig03}. At large values of the coupling $t_{MK}$, the two MFs at the two ends (away from the 
 Josephson junction) are close to zero energy and they exhibit almost perfect Andreev reflection. Along the arcs outside the energy range 
 $(-0.5\De,0.5\De)$, local Andreev reflection is weak but the nonlocal transport is enhanced. A threshold hopping strength $t_{MK}$ is 
 necessary for the nonlocal transport to dominate.  We see that CAR is enhanced over ET along the arc in positive bias while ET is 
 enhanced over CAR on the arc in negative bias.  
 
 Further, we find that an asymmetry in the hopping amplitudes from the NM leads to the Josephson junction leads to rich features in 
 the conductance results. We find that $0<t_L<t_R\le t$ enhances the nonlocal transport by the following mechanism. An electron enters the Josephson
 junction at the resonant energies by resonant tunneling from $N_L$ despite a weak coupling $t_L$. The electron gets converted into BdG
 quasiparticle in the Josephson junction.  Now, having $t_R>t_L$ will increase the transparency of the BdG quasiparticle into $N_R$ than that 
 into $N_L$. So, the transmission (either as an electron or as a hole) onto $N_R$ is enhanced, there by aiding nonlocal transport. 
 A similar mechanism explains why local transport is enhanced in the limit $0<t_R<t_L\le t$ when a bias is applied from left to 
 right. 
 
 There are four MFs in the Josephson junction that is made of two KCs. The two MFs in single KC are coupled since $\De\neq t$. 
 Further, the MFs in the two KCs are coupled due to a finite $t_{MK}$. So, the four Dirac fermion states in the energy range 
 $(-1.5\De,1.5\De)$ in Fig.~\ref{fig01} are formed by the hybridization of four MFs that are nonlocal. Hence, the negative
 transconductance at energies of ABSs is  a sure~sign of nonlocality of constituent MFs. 

\section{Discussion and Conclusion}
A stark contrast between Josephson junctions made of s-wave superconductors and those made of p-wave superconductors hosting 
MFs is that p-wave superconductors exhibit $4\pi$ Josephson effect whereas s-wave superconductors exhibit $2\pi$ Josephson effect. 
This happens because of crossing of ABS energy levels at $\phi=\pi$ in a Josephson junction made out of p-wave superconductor hosting 
isolated MFs at its ends. Of the two ABS levels formed by hybridization of MFs, the one with lower energy is occupied for 
$0\le\phi\le\pi$ and $3\pi\le\phi\le 4\pi$ while the the ABS with higher energy is occupied for $\pi\le\phi\le 3\pi$. This is 
true if we are looking purely at the Josephson current between two p-wave superconductors. But when normal metal leads are 
connected to the ends of Josephson junction, the occupancy of the lower and the higher ABS levels in the isolated Josephson junction
does not affect the current driven in the normal metal leads by the applied bias. Hence the $4\pi$ periodicity is not manifest in 
the local conductance and transconductance of the  proposed setup. 

We have seen that the Josephson junction of Kitaev chains connected to normal metal leads can be used to enhance crossed Andreev
reflection over electron tunneling. This means the setup can be used to probe the nonlocality of the constituent Majorana
fermions. The proposal presented in this work is different from our previous proposal~\cite{soori19trans} in many respects. 
Firstly, the peaks in local conductance and the valleys and peaks in the transconductance match with the energy levels of the isolated 
Josephson junction. This indicates that the negative transconductance in the energy gap is purely due to the constituent Majorana
fermions in contrast to the previous proposal where the negative transconductance was due to the combined effect of Majorana fermions
and the subgap Andreev states from the continuum band of Kitaev ladder~\cite{soori19trans}.  Secondly, we find that asymmetry in couplings
of the Josephson junction to the normal metal can enhance nonlocal transport. Thirdly, since the Josephson junctions between 
p-wave superconductors hosting Majorana fermions have already been fabricated~\cite{rokhi2012} this proposal can be tested 
experimentally with present technology.

Experimentally, p-wave superconducting quantum wires hosting Majorana fermions at the ends have been 
fabricated~\cite{Das2012,albrecht2016,Zhang2018,aguado17}. We propose to fabricate two p-wave superconducting quantum  wires connected 
end to end via a point contact. The superconductivity in such quantum wires is induced by a bulk s-wave superconductor in proximity. 
The Josephson phase difference can be induced by forming a SQUID loop between the two bulk superconductors that are in proximity to the 
quantum wires. The two p-wave superconducting quantum wires must be connected to two normal metal leads at the two ends. In a setup designed
so, the conditions to observe negative transconductance are: (i)~the parameters for each KC should be chosen so that the quantum wire is
in topological phase, (ii)~the Josephson junction should be transparent enough to allow for
hybridization between the two MFs, (iii)~the length of each of the p-wave superconducting quantum  wire must be optimal which is decided
by the strength of the induced superconducting pairing, long enough for the MFs to be formed within the gap and short enough so that the 
two MFs at the ends of each KC hybridize. 
\acknowledgements
The author thanks DST-INSPIRE Faculty Award (Faculty Reg. No.~:~IFA17-PH190)
for financial support. The author thanks Diptiman Sen for comments on the manuscript and Karthik V Raman for discussions
on experimental feasibility.

\bibliography{ref_mfjo}

\begin{thebibliography}{41}%
\makeatletter
\providecommand \@ifxundefined [1]{%
 \@ifx{#1\undefined}
}%
\providecommand \@ifnum [1]{%
 \ifnum #1\expandafter \@firstoftwo
 \else \expandafter \@secondoftwo
 \fi
}%
\providecommand \@ifx [1]{%
 \ifx #1\expandafter \@firstoftwo
 \else \expandafter \@secondoftwo
 \fi
}%
\providecommand \natexlab [1]{#1}%
\providecommand \enquote  [1]{``#1''}%
\providecommand \bibnamefont  [1]{#1}%
\providecommand \bibfnamefont [1]{#1}%
\providecommand \citenamefont [1]{#1}%
\providecommand \href@noop [0]{\@secondoftwo}%
\providecommand \href [0]{\begingroup \@sanitize@url \@href}%
\providecommand \@href[1]{\@@startlink{#1}\@@href}%
\providecommand \@@href[1]{\endgroup#1\@@endlink}%
\providecommand \@sanitize@url [0]{\catcode `\\12\catcode `\$12\catcode
  `\&12\catcode `\#12\catcode `\^12\catcode `\_12\catcode `\%12\relax}%
\providecommand \@@startlink[1]{}%
\providecommand \@@endlink[0]{}%
\providecommand \url  [0]{\begingroup\@sanitize@url \@url }%
\providecommand \@url [1]{\endgroup\@href {#1}{\urlprefix }}%
\providecommand \urlprefix  [0]{URL }%
\providecommand \Eprint [0]{\href }%
\providecommand \doibase [0]{http://dx.doi.org/}%
\providecommand \selectlanguage [0]{\@gobble}%
\providecommand \bibinfo  [0]{\@secondoftwo}%
\providecommand \bibfield  [0]{\@secondoftwo}%
\providecommand \translation [1]{[#1]}%
\providecommand \BibitemOpen [0]{}%
\providecommand \bibitemStop [0]{}%
\providecommand \bibitemNoStop [0]{.\EOS\space}%
\providecommand \EOS [0]{\spacefactor3000\relax}%
\providecommand \BibitemShut  [1]{\csname bibitem#1\endcsname}%
\let\auto@bib@innerbib\@empty
\bibitem [{\citenamefont {Kitaev}(2001)}]{kitaev2001unpaired}%
  \BibitemOpen
  \bibfield  {author} {\bibinfo {author} {\bibfnamefont {A.~Y.}\ \bibnamefont
  {Kitaev}},\ }\bibfield  {title} {\enquote {\bibinfo {title} {Unpaired
  majorana fermions in quantum wires},}\ }\href {\doibase
  10.1070/1063-7869/44/10s/s29} {\bibfield  {journal} {\bibinfo  {journal}
  {Phys.-Usp.}\ }\textbf {\bibinfo {volume} {44}},\ \bibinfo {pages} {131}
  (\bibinfo {year} {2001})}\BibitemShut {NoStop}%
\bibitem [{\citenamefont {Nayak}\ \emph {et~al.}(2008)\citenamefont {Nayak},
  \citenamefont {Simon}, \citenamefont {Stern}, \citenamefont {Freedman},\ and\
  \citenamefont {Das~Sarma}}]{nayak08}%
  \BibitemOpen
  \bibfield  {author} {\bibinfo {author} {\bibfnamefont {C.}~\bibnamefont
  {Nayak}}, \bibinfo {author} {\bibfnamefont {S.~H.}\ \bibnamefont {Simon}},
  \bibinfo {author} {\bibfnamefont {A.}~\bibnamefont {Stern}}, \bibinfo
  {author} {\bibfnamefont {M.}~\bibnamefont {Freedman}}, \ and\ \bibinfo
  {author} {\bibfnamefont {S.}~\bibnamefont {Das~Sarma}},\ }\bibfield  {title}
  {\enquote {\bibinfo {title} {Non-abelian anyons and topological quantum
  computation},}\ }\href {\doibase 10.1103/RevModPhys.80.1083} {\bibfield
  {journal} {\bibinfo  {journal} {Rev. Mod. Phys.}\ }\textbf {\bibinfo {volume}
  {80}},\ \bibinfo {pages} {1083--1159} (\bibinfo {year} {2008})}\BibitemShut
  {NoStop}%
\bibitem [{\citenamefont {Lieb}\ \emph {et~al.}(1961)\citenamefont {Lieb},
  \citenamefont {Schultz},\ and\ \citenamefont {Mattis}}]{lieb1961}%
  \BibitemOpen
  \bibfield  {author} {\bibinfo {author} {\bibfnamefont {E.}~\bibnamefont
  {Lieb}}, \bibinfo {author} {\bibfnamefont {T.}~\bibnamefont {Schultz}}, \
  and\ \bibinfo {author} {\bibfnamefont {D.}~\bibnamefont {Mattis}},\
  }\bibfield  {title} {\enquote {\bibinfo {title} {Two soluble models of an
  antiferromagnetic chain},}\ }\href {\doibase
  https://doi.org/10.1016/0003-4916(61)90115-4} {\bibfield  {journal} {\bibinfo
   {journal} {Annals of Physics}\ }\textbf {\bibinfo {volume} {16}},\ \bibinfo
  {pages} {407 -- 466} (\bibinfo {year} {1961})}\BibitemShut {NoStop}%
\bibitem [{\citenamefont {Lutchyn}\ \emph {et~al.}(2010)\citenamefont
  {Lutchyn}, \citenamefont {Sau},\ and\ \citenamefont
  {Das~Sarma}}]{lutchyn2010majorana}%
  \BibitemOpen
  \bibfield  {author} {\bibinfo {author} {\bibfnamefont {R.~M.}\ \bibnamefont
  {Lutchyn}}, \bibinfo {author} {\bibfnamefont {J.~D.}\ \bibnamefont {Sau}}, \
  and\ \bibinfo {author} {\bibfnamefont {S.}~\bibnamefont {Das~Sarma}},\
  }\bibfield  {title} {\enquote {\bibinfo {title} {Majorana fermions and a
  topological phase transition in semiconductor-superconductor
  heterostructures},}\ }\href {\doibase 10.1103/PhysRevLett.105.077001}
  {\bibfield  {journal} {\bibinfo  {journal} {Phys. Rev. Lett.}\ }\textbf
  {\bibinfo {volume} {105}},\ \bibinfo {pages} {077001} (\bibinfo {year}
  {2010})}\BibitemShut {NoStop}%
\bibitem [{\citenamefont {Oreg}\ \emph {et~al.}(2010)\citenamefont {Oreg},
  \citenamefont {Refael},\ and\ \citenamefont {von Oppen}}]{oreg2010helical}%
  \BibitemOpen
  \bibfield  {author} {\bibinfo {author} {\bibfnamefont {Y.}~\bibnamefont
  {Oreg}}, \bibinfo {author} {\bibfnamefont {G.}~\bibnamefont {Refael}}, \ and\
  \bibinfo {author} {\bibfnamefont {F.}~\bibnamefont {von Oppen}},\ }\bibfield
  {title} {\enquote {\bibinfo {title} {Helical liquids and majorana bound
  states in quantum wires},}\ }\href {\doibase 10.1103/PhysRevLett.105.177002}
  {\bibfield  {journal} {\bibinfo  {journal} {Phys. Rev. Lett.}\ }\textbf
  {\bibinfo {volume} {105}},\ \bibinfo {pages} {177002} (\bibinfo {year}
  {2010})}\BibitemShut {NoStop}%
\bibitem [{\citenamefont {Mourik}\ \emph {et~al.}(2012)\citenamefont {Mourik},
  \citenamefont {Zuo}, \citenamefont {Frolov}, \citenamefont {Plissard},
  \citenamefont {Bakkers},\ and\ \citenamefont {Kouwenhoven}}]{mourik2012}%
  \BibitemOpen
  \bibfield  {author} {\bibinfo {author} {\bibfnamefont {V.}~\bibnamefont
  {Mourik}}, \bibinfo {author} {\bibfnamefont {K.}~\bibnamefont {Zuo}},
  \bibinfo {author} {\bibfnamefont {S.~M.}\ \bibnamefont {Frolov}}, \bibinfo
  {author} {\bibfnamefont {S.R.}\ \bibnamefont {Plissard}}, \bibinfo {author}
  {\bibfnamefont {E.~P. A.~M.}\ \bibnamefont {Bakkers}}, \ and\ \bibinfo
  {author} {\bibfnamefont {L.~P.}\ \bibnamefont {Kouwenhoven}},\ }\bibfield
  {title} {\enquote {\bibinfo {title} {Signatures of majorana fermions in
  hybrid superconductor-semiconductor nanowire devices},}\ }\href {\doibase
  10.1126/science.1222360} {\bibfield  {journal} {\bibinfo  {journal}
  {Science}\ }\textbf {\bibinfo {volume} {336}},\ \bibinfo {pages} {1003--1007}
  (\bibinfo {year} {2012})}\BibitemShut {NoStop}%
\bibitem [{\citenamefont {Das}\ \emph {et~al.}(2012)\citenamefont {Das},
  \citenamefont {Ronen}, \citenamefont {Most}, \citenamefont {Oreg},
  \citenamefont {Heiblum},\ and\ \citenamefont {Shtrikman}}]{Das2012}%
  \BibitemOpen
  \bibfield  {author} {\bibinfo {author} {\bibfnamefont {A.}~\bibnamefont
  {Das}}, \bibinfo {author} {\bibfnamefont {Y.}~\bibnamefont {Ronen}}, \bibinfo
  {author} {\bibfnamefont {Y.}~\bibnamefont {Most}}, \bibinfo {author}
  {\bibfnamefont {Y.}~\bibnamefont {Oreg}}, \bibinfo {author} {\bibfnamefont
  {M.}~\bibnamefont {Heiblum}}, \ and\ \bibinfo {author} {\bibfnamefont
  {H.}~\bibnamefont {Shtrikman}},\ }\bibfield  {title} {\enquote {\bibinfo
  {title} {Zero-bias peaks and splitting in an al-inas nanowire topological
  superconductor as a signature of majorana fermions},}\ }\href {\doibase
  10.1038/nphys2479} {\bibfield  {journal} {\bibinfo  {journal} {Nat. Phys.}\
  }\textbf {\bibinfo {volume} {8}},\ \bibinfo {pages} {887} (\bibinfo {year}
  {2012})}\BibitemShut {NoStop}%
\bibitem [{\citenamefont {Albrecht}\ \emph {et~al.}(2016)\citenamefont
  {Albrecht}, \citenamefont {Higginbotham}, \citenamefont {Madsen},
  \citenamefont {Kuemmeth}, \citenamefont {Jespersen}, \citenamefont
  {Nyg{\aa}rd}, \citenamefont {Krogstrup},\ and\ \citenamefont
  {Marcus}}]{albrecht2016}%
  \BibitemOpen
  \bibfield  {author} {\bibinfo {author} {\bibfnamefont {S.~M.}\ \bibnamefont
  {Albrecht}}, \bibinfo {author} {\bibfnamefont {A.~P.}\ \bibnamefont
  {Higginbotham}}, \bibinfo {author} {\bibfnamefont {M.}~\bibnamefont
  {Madsen}}, \bibinfo {author} {\bibfnamefont {F.}~\bibnamefont {Kuemmeth}},
  \bibinfo {author} {\bibfnamefont {T.~S.}\ \bibnamefont {Jespersen}}, \bibinfo
  {author} {\bibfnamefont {J.}~\bibnamefont {Nyg{\aa}rd}}, \bibinfo {author}
  {\bibfnamefont {P.}~\bibnamefont {Krogstrup}}, \ and\ \bibinfo {author}
  {\bibfnamefont {C.~M.}\ \bibnamefont {Marcus}},\ }\bibfield  {title}
  {\enquote {\bibinfo {title} {Exponential protection of zero modes in majorana
  islands},}\ }\href {\doibase 10.1038/nature17162} {\bibfield  {journal}
  {\bibinfo  {journal} {Nature}\ }\textbf {\bibinfo {volume} {531}},\ \bibinfo
  {pages} {206} (\bibinfo {year} {2016})}\BibitemShut {NoStop}%
\bibitem [{\citenamefont {Zhang}\ \emph {et~al.}(2018)\citenamefont {Zhang},
  \citenamefont {Liu}, \citenamefont {Gazibegovic}, \citenamefont {Xu},
  \citenamefont {Logan}, \citenamefont {Wang}, \citenamefont {van Loo},
  \citenamefont {Bommer}, \citenamefont {de~Moor}, \citenamefont {Car},
  \citenamefont {Op~het Veld}, \citenamefont {van Veldhoven}, \citenamefont
  {Koelling}, \citenamefont {Verheijen}, \citenamefont {Pendharkar},
  \citenamefont {Pennachio}, \citenamefont {Shojaei}, \citenamefont {Lee},
  \citenamefont {Palmstr{\o}m}, \citenamefont {Bakkers}, \citenamefont
  {Sarma},\ and\ \citenamefont {Kouwenhoven}}]{Zhang2018}%
  \BibitemOpen
  \bibfield  {author} {\bibinfo {author} {\bibfnamefont {H.}~\bibnamefont
  {Zhang}}, \bibinfo {author} {\bibfnamefont {C.-X.}\ \bibnamefont {Liu}},
  \bibinfo {author} {\bibfnamefont {S.}~\bibnamefont {Gazibegovic}}, \bibinfo
  {author} {\bibfnamefont {D.}~\bibnamefont {Xu}}, \bibinfo {author}
  {\bibfnamefont {J.~A.}\ \bibnamefont {Logan}}, \bibinfo {author}
  {\bibfnamefont {G.}~\bibnamefont {Wang}}, \bibinfo {author} {\bibfnamefont
  {N.}~\bibnamefont {van Loo}}, \bibinfo {author} {\bibfnamefont {J.~D.~S.}\
  \bibnamefont {Bommer}}, \bibinfo {author} {\bibfnamefont {M.~W.~A.}\
  \bibnamefont {de~Moor}}, \bibinfo {author} {\bibfnamefont {D.}~\bibnamefont
  {Car}}, \bibinfo {author} {\bibfnamefont {R.~L.~M.}\ \bibnamefont {Op~het
  Veld}}, \bibinfo {author} {\bibfnamefont {P.~J.}\ \bibnamefont {van
  Veldhoven}}, \bibinfo {author} {\bibfnamefont {S.}~\bibnamefont {Koelling}},
  \bibinfo {author} {\bibfnamefont {M.~A.}\ \bibnamefont {Verheijen}}, \bibinfo
  {author} {\bibfnamefont {M.}~\bibnamefont {Pendharkar}}, \bibinfo {author}
  {\bibfnamefont {D.~J.}\ \bibnamefont {Pennachio}}, \bibinfo {author}
  {\bibfnamefont {B.}~\bibnamefont {Shojaei}}, \bibinfo {author} {\bibfnamefont
  {J.~S.}\ \bibnamefont {Lee}}, \bibinfo {author} {\bibfnamefont {C.~J.}\
  \bibnamefont {Palmstr{\o}m}}, \bibinfo {author} {\bibfnamefont {E.~P. A.~M.}\
  \bibnamefont {Bakkers}}, \bibinfo {author} {\bibfnamefont {S.~D.}\
  \bibnamefont {Sarma}}, \ and\ \bibinfo {author} {\bibfnamefont {L.~P.}\
  \bibnamefont {Kouwenhoven}},\ }\bibfield  {title} {\enquote {\bibinfo {title}
  {Quantized majorana conductance},}\ }\href {\doibase 10.1038/nature26142}
  {\bibfield  {journal} {\bibinfo  {journal} {Nature}\ }\textbf {\bibinfo
  {volume} {556}},\ \bibinfo {pages} {74} (\bibinfo {year} {2018})}\BibitemShut
  {NoStop}%
\bibitem [{\citenamefont {Aguado}(2017)}]{aguado17}%
  \BibitemOpen
  \bibfield  {author} {\bibinfo {author} {\bibfnamefont {R.}~\bibnamefont
  {Aguado}},\ }\bibfield  {title} {\enquote {\bibinfo {title} {Majorana
  quasiparticles in condensed matter},}\ }\href@noop {} {\bibfield  {journal}
  {\bibinfo  {journal} {La Rivista del Nuovo Cimento}\ }\textbf {\bibinfo
  {volume} {40}},\ \bibinfo {pages} {523} (\bibinfo {year} {2017})}\BibitemShut
  {NoStop}%
\bibitem [{\citenamefont {Liu}\ \emph {et~al.}(2017)\citenamefont {Liu},
  \citenamefont {Song}, \citenamefont {Sun},\ and\ \citenamefont
  {Xie}}]{liu17}%
  \BibitemOpen
  \bibfield  {author} {\bibinfo {author} {\bibfnamefont {J}~\bibnamefont
  {Liu}}, \bibinfo {author} {\bibfnamefont {J.}~\bibnamefont {Song}}, \bibinfo
  {author} {\bibfnamefont {Q.-F.}\ \bibnamefont {Sun}}, \ and\ \bibinfo
  {author} {\bibfnamefont {X.~C.}\ \bibnamefont {Xie}},\ }\bibfield  {title}
  {\enquote {\bibinfo {title} {Even-odd interference effect in a topological
  superconducting wire},}\ }\href {\doibase 10.1103/PhysRevB.96.195307}
  {\bibfield  {journal} {\bibinfo  {journal} {Phys. Rev. B}\ }\textbf {\bibinfo
  {volume} {96}},\ \bibinfo {pages} {195307} (\bibinfo {year}
  {2017})}\BibitemShut {NoStop}%
\bibitem [{\citenamefont {Nilsson}\ \emph {et~al.}(2008)\citenamefont
  {Nilsson}, \citenamefont {Akhmerov},\ and\ \citenamefont
  {Beenakker}}]{nilsson2008}%
  \BibitemOpen
  \bibfield  {author} {\bibinfo {author} {\bibfnamefont {J.}~\bibnamefont
  {Nilsson}}, \bibinfo {author} {\bibfnamefont {A.R.}\ \bibnamefont
  {Akhmerov}}, \ and\ \bibinfo {author} {\bibfnamefont {C.~W.~J.}\ \bibnamefont
  {Beenakker}},\ }\bibfield  {title} {\enquote {\bibinfo {title} {Splitting of
  a cooper pair by a pair of majorana bound states},}\ }\href {\doibase
  10.1103/PhysRevLett.101.120403} {\bibfield  {journal} {\bibinfo  {journal}
  {Phys. Rev. Lett.}\ }\textbf {\bibinfo {volume} {101}},\ \bibinfo {pages}
  {120403} (\bibinfo {year} {2008})}\BibitemShut {NoStop}%
\bibitem [{\citenamefont {L\"u}\ \emph {et~al.}(2012)\citenamefont {L\"u},
  \citenamefont {Lu},\ and\ \citenamefont {Shen}}]{Lu12}%
  \BibitemOpen
  \bibfield  {author} {\bibinfo {author} {\bibfnamefont {H.-F.}\ \bibnamefont
  {L\"u}}, \bibinfo {author} {\bibfnamefont {H.-Z.}\ \bibnamefont {Lu}}, \ and\
  \bibinfo {author} {\bibfnamefont {S.-Q.}\ \bibnamefont {Shen}},\ }\bibfield
  {title} {\enquote {\bibinfo {title} {Nonlocal noise cross correlation
  mediated by entangled majorana fermions},}\ }\href {\doibase
  10.1103/PhysRevB.86.075318} {\bibfield  {journal} {\bibinfo  {journal} {Phys.
  Rev. B}\ }\textbf {\bibinfo {volume} {86}},\ \bibinfo {pages} {075318}
  (\bibinfo {year} {2012})}\BibitemShut {NoStop}%
\bibitem [{\citenamefont {Liu}\ \emph {et~al.}(2013)\citenamefont {Liu},
  \citenamefont {Zhang},\ and\ \citenamefont {Law}}]{Liu13}%
  \BibitemOpen
  \bibfield  {author} {\bibinfo {author} {\bibfnamefont {J.}~\bibnamefont
  {Liu}}, \bibinfo {author} {\bibfnamefont {F.-C.}\ \bibnamefont {Zhang}}, \
  and\ \bibinfo {author} {\bibfnamefont {K.~T.}\ \bibnamefont {Law}},\
  }\bibfield  {title} {\enquote {\bibinfo {title} {Majorana fermion induced
  nonlocal current correlations in spin-orbit coupled superconducting wires},}\
  }\href {\doibase 10.1103/PhysRevB.88.064509} {\bibfield  {journal} {\bibinfo
  {journal} {Phys. Rev. B}\ }\textbf {\bibinfo {volume} {88}},\ \bibinfo
  {pages} {064509} (\bibinfo {year} {2013})}\BibitemShut {NoStop}%
\bibitem [{\citenamefont {Zocher}\ and\ \citenamefont
  {Rosenow}(2013)}]{Zocher13}%
  \BibitemOpen
  \bibfield  {author} {\bibinfo {author} {\bibfnamefont {B.}~\bibnamefont
  {Zocher}}\ and\ \bibinfo {author} {\bibfnamefont {B.}~\bibnamefont
  {Rosenow}},\ }\bibfield  {title} {\enquote {\bibinfo {title} {Modulation of
  majorana-induced current cross-correlations by quantum dots},}\ }\href
  {\doibase 10.1103/PhysRevLett.111.036802} {\bibfield  {journal} {\bibinfo
  {journal} {Phys. Rev. Lett.}\ }\textbf {\bibinfo {volume} {111}},\ \bibinfo
  {pages} {036802} (\bibinfo {year} {2013})}\BibitemShut {NoStop}%
\bibitem [{\citenamefont {Prada}\ \emph {et~al.}(2017)\citenamefont {Prada},
  \citenamefont {Aguado},\ and\ \citenamefont {San-Jose}}]{Prada17}%
  \BibitemOpen
  \bibfield  {author} {\bibinfo {author} {\bibfnamefont {E.}~\bibnamefont
  {Prada}}, \bibinfo {author} {\bibfnamefont {R.}~\bibnamefont {Aguado}}, \
  and\ \bibinfo {author} {\bibfnamefont {P.}~\bibnamefont {San-Jose}},\
  }\bibfield  {title} {\enquote {\bibinfo {title} {Measuring majorana
  nonlocality and spin structure with a quantum dot},}\ }\href {\doibase
  10.1103/PhysRevB.96.085418} {\bibfield  {journal} {\bibinfo  {journal} {Phys.
  Rev. B}\ }\textbf {\bibinfo {volume} {96}},\ \bibinfo {pages} {085418}
  (\bibinfo {year} {2017})}\BibitemShut {NoStop}%
\bibitem [{\citenamefont {Nehra}\ \emph {et~al.}(2019)\citenamefont {Nehra},
  \citenamefont {Bhakuni}, \citenamefont {Sharma},\ and\ \citenamefont
  {Soori}}]{nehra19}%
  \BibitemOpen
  \bibfield  {author} {\bibinfo {author} {\bibfnamefont {R.}~\bibnamefont
  {Nehra}}, \bibinfo {author} {\bibfnamefont {D.~S.}\ \bibnamefont {Bhakuni}},
  \bibinfo {author} {\bibfnamefont {A.}~\bibnamefont {Sharma}}, \ and\ \bibinfo
  {author} {\bibfnamefont {A.}~\bibnamefont {Soori}},\ }\bibfield  {title}
  {\enquote {\bibinfo {title} {Enhancement of crossed andreev reflection in a
  kitaev ladder connected to normal metal leads},}\ }\href {\doibase
  10.1088/1361-648x/ab2403} {\bibfield  {journal} {\bibinfo  {journal} {J.
  Phys.: Condens. Matter}\ }\textbf {\bibinfo {volume} {31}},\ \bibinfo {pages}
  {345304} (\bibinfo {year} {2019})}\BibitemShut {NoStop}%
\bibitem [{\citenamefont {Soori}(2019)}]{soori19trans}%
  \BibitemOpen
  \bibfield  {author} {\bibinfo {author} {\bibfnamefont {A.}~\bibnamefont
  {Soori}},\ }\bibfield  {title} {\enquote {\bibinfo {title} {Transconductance
  as a probe of nonlocality of majorana fermions},}\ }\href {\doibase
  10.1088/1361-648x/ab3f73} {\bibfield  {journal} {\bibinfo  {journal} {J.
  Phys.: Condens. Matter}\ }\textbf {\bibinfo {volume} {31}},\ \bibinfo {pages}
  {505301} (\bibinfo {year} {2019})}\BibitemShut {NoStop}%
\bibitem [{\citenamefont {Fu}(2010)}]{fu10}%
  \BibitemOpen
  \bibfield  {author} {\bibinfo {author} {\bibfnamefont {L.}~\bibnamefont
  {Fu}},\ }\bibfield  {title} {\enquote {\bibinfo {title} {Electron
  teleportation via majorana bound states in a mesoscopic superconductor},}\
  }\href {\doibase 10.1103/PhysRevLett.104.056402} {\bibfield  {journal}
  {\bibinfo  {journal} {Phys. Rev. Lett.}\ }\textbf {\bibinfo {volume} {104}},\
  \bibinfo {pages} {056402} (\bibinfo {year} {2010})}\BibitemShut {NoStop}%
\bibitem [{\citenamefont {Zazunov}\ \emph {et~al.}(2011)\citenamefont
  {Zazunov}, \citenamefont {Yeyati},\ and\ \citenamefont {Egger}}]{zazunov11}%
  \BibitemOpen
  \bibfield  {author} {\bibinfo {author} {\bibfnamefont {A.}~\bibnamefont
  {Zazunov}}, \bibinfo {author} {\bibfnamefont {A.~Levy}\ \bibnamefont
  {Yeyati}}, \ and\ \bibinfo {author} {\bibfnamefont {R.}~\bibnamefont
  {Egger}},\ }\bibfield  {title} {\enquote {\bibinfo {title} {Coulomb blockade
  of majorana-fermion-induced transport},}\ }\href {\doibase
  10.1103/PhysRevB.84.165440} {\bibfield  {journal} {\bibinfo  {journal} {Phys.
  Rev. B}\ }\textbf {\bibinfo {volume} {84}},\ \bibinfo {pages} {165440}
  (\bibinfo {year} {2011})}\BibitemShut {NoStop}%
\bibitem [{\citenamefont {H\"utzen}\ \emph {et~al.}(2012)\citenamefont
  {H\"utzen}, \citenamefont {Zazunov}, \citenamefont {Braunecker},
  \citenamefont {Yeyati},\ and\ \citenamefont {Egger}}]{hutzen12}%
  \BibitemOpen
  \bibfield  {author} {\bibinfo {author} {\bibfnamefont {R.}~\bibnamefont
  {H\"utzen}}, \bibinfo {author} {\bibfnamefont {A.}~\bibnamefont {Zazunov}},
  \bibinfo {author} {\bibfnamefont {B.}~\bibnamefont {Braunecker}}, \bibinfo
  {author} {\bibfnamefont {A.~Levy}\ \bibnamefont {Yeyati}}, \ and\ \bibinfo
  {author} {\bibfnamefont {R.}~\bibnamefont {Egger}},\ }\bibfield  {title}
  {\enquote {\bibinfo {title} {Majorana single-charge transistor},}\ }\href
  {\doibase 10.1103/PhysRevLett.109.166403} {\bibfield  {journal} {\bibinfo
  {journal} {Phys. Rev. Lett.}\ }\textbf {\bibinfo {volume} {109}},\ \bibinfo
  {pages} {166403} (\bibinfo {year} {2012})}\BibitemShut {NoStop}%
\bibitem [{\citenamefont {Zazunov}\ and\ \citenamefont
  {Egger}(2012)}]{zazunov12}%
  \BibitemOpen
  \bibfield  {author} {\bibinfo {author} {\bibfnamefont {A.}~\bibnamefont
  {Zazunov}}\ and\ \bibinfo {author} {\bibfnamefont {R.}~\bibnamefont
  {Egger}},\ }\bibfield  {title} {\enquote {\bibinfo {title} {Supercurrent
  blockade in josephson junctions with a majorana wire},}\ }\href {\doibase
  10.1103/PhysRevB.85.104514} {\bibfield  {journal} {\bibinfo  {journal} {Phys.
  Rev. B}\ }\textbf {\bibinfo {volume} {85}},\ \bibinfo {pages} {104514}
  (\bibinfo {year} {2012})}\BibitemShut {NoStop}%
\bibitem [{\citenamefont {Wang}\ \emph {et~al.}(2013)\citenamefont {Wang},
  \citenamefont {Hu}, \citenamefont {Liang},\ and\ \citenamefont
  {Hu}}]{wang13}%
  \BibitemOpen
  \bibfield  {author} {\bibinfo {author} {\bibfnamefont {Z.}~\bibnamefont
  {Wang}}, \bibinfo {author} {\bibfnamefont {X-Y.}\ \bibnamefont {Hu}},
  \bibinfo {author} {\bibfnamefont {Q-F.}\ \bibnamefont {Liang}}, \ and\
  \bibinfo {author} {\bibfnamefont {X.}~\bibnamefont {Hu}},\ }\bibfield
  {title} {\enquote {\bibinfo {title} {Detecting majorana fermions by nonlocal
  entanglement between quantum dots},}\ }\href {\doibase
  10.1103/PhysRevB.87.214513} {\bibfield  {journal} {\bibinfo  {journal} {Phys.
  Rev. B}\ }\textbf {\bibinfo {volume} {87}},\ \bibinfo {pages} {214513}
  (\bibinfo {year} {2013})}\BibitemShut {NoStop}%
\bibitem [{\citenamefont {Sau}\ \emph {et~al.}(2015)\citenamefont {Sau},
  \citenamefont {Swingle},\ and\ \citenamefont {Tewari}}]{sau15}%
  \BibitemOpen
  \bibfield  {author} {\bibinfo {author} {\bibfnamefont {J.~D.}\ \bibnamefont
  {Sau}}, \bibinfo {author} {\bibfnamefont {B.}~\bibnamefont {Swingle}}, \ and\
  \bibinfo {author} {\bibfnamefont {S.}~\bibnamefont {Tewari}},\ }\bibfield
  {title} {\enquote {\bibinfo {title} {Proposal to probe quantum nonlocality of
  majorana fermions in tunneling experiments},}\ }\href {\doibase
  10.1103/PhysRevB.92.020511} {\bibfield  {journal} {\bibinfo  {journal} {Phys.
  Rev. B}\ }\textbf {\bibinfo {volume} {92}},\ \bibinfo {pages} {020511}
  (\bibinfo {year} {2015})}\BibitemShut {NoStop}%
\bibitem [{\citenamefont {B\'eri}\ and\ \citenamefont {Cooper}(2012)}]{beri12}%
  \BibitemOpen
  \bibfield  {author} {\bibinfo {author} {\bibfnamefont {B.}~\bibnamefont
  {B\'eri}}\ and\ \bibinfo {author} {\bibfnamefont {N.~R.}\ \bibnamefont
  {Cooper}},\ }\bibfield  {title} {\enquote {\bibinfo {title} {Topological
  kondo effect with majorana fermions},}\ }\href {\doibase
  10.1103/PhysRevLett.109.156803} {\bibfield  {journal} {\bibinfo  {journal}
  {Phys. Rev. Lett.}\ }\textbf {\bibinfo {volume} {109}},\ \bibinfo {pages}
  {156803} (\bibinfo {year} {2012})}\BibitemShut {NoStop}%
\bibitem [{\citenamefont {Altland}\ and\ \citenamefont
  {Egger}(2013)}]{altland13}%
  \BibitemOpen
  \bibfield  {author} {\bibinfo {author} {\bibfnamefont {A.}~\bibnamefont
  {Altland}}\ and\ \bibinfo {author} {\bibfnamefont {R.}~\bibnamefont
  {Egger}},\ }\bibfield  {title} {\enquote {\bibinfo {title} {Multiterminal
  coulomb-majorana junction},}\ }\href {\doibase
  10.1103/PhysRevLett.110.196401} {\bibfield  {journal} {\bibinfo  {journal}
  {Phys. Rev. Lett.}\ }\textbf {\bibinfo {volume} {110}},\ \bibinfo {pages}
  {196401} (\bibinfo {year} {2013})}\BibitemShut {NoStop}%
\bibitem [{\citenamefont {B\'eri}(2013)}]{beri13}%
  \BibitemOpen
  \bibfield  {author} {\bibinfo {author} {\bibfnamefont {B.}~\bibnamefont
  {B\'eri}},\ }\bibfield  {title} {\enquote {\bibinfo {title} {Majorana-klein
  hybridization in topological superconductor junctions},}\ }\href {\doibase
  10.1103/PhysRevLett.110.216803} {\bibfield  {journal} {\bibinfo  {journal}
  {Phys. Rev. Lett.}\ }\textbf {\bibinfo {volume} {110}},\ \bibinfo {pages}
  {216803} (\bibinfo {year} {2013})}\BibitemShut {NoStop}%
\bibitem [{\citenamefont {Altland}\ \emph {et~al.}(2014)\citenamefont
  {Altland}, \citenamefont {B\'eri}, \citenamefont {Egger},\ and\ \citenamefont
  {Tsvelik}}]{altland14}%
  \BibitemOpen
  \bibfield  {author} {\bibinfo {author} {\bibfnamefont {A.}~\bibnamefont
  {Altland}}, \bibinfo {author} {\bibfnamefont {B.}~\bibnamefont {B\'eri}},
  \bibinfo {author} {\bibfnamefont {R.}~\bibnamefont {Egger}}, \ and\ \bibinfo
  {author} {\bibfnamefont {A.~M.}\ \bibnamefont {Tsvelik}},\ }\bibfield
  {title} {\enquote {\bibinfo {title} {Multichannel kondo impurity dynamics in
  a majorana device},}\ }\href {\doibase 10.1103/PhysRevLett.113.076401}
  {\bibfield  {journal} {\bibinfo  {journal} {Phys. Rev. Lett.}\ }\textbf
  {\bibinfo {volume} {113}},\ \bibinfo {pages} {076401} (\bibinfo {year}
  {2014})}\BibitemShut {NoStop}%
\bibitem [{\citenamefont {van Beek}\ and\ \citenamefont
  {Braunecker}(2016)}]{vanBeek16}%
  \BibitemOpen
  \bibfield  {author} {\bibinfo {author} {\bibfnamefont {I.~J.}\ \bibnamefont
  {van Beek}}\ and\ \bibinfo {author} {\bibfnamefont {B.}~\bibnamefont
  {Braunecker}},\ }\bibfield  {title} {\enquote {\bibinfo {title} {Non-kondo
  many-body physics in a majorana-based kondo type system},}\ }\href {\doibase
  10.1103/PhysRevB.94.115416} {\bibfield  {journal} {\bibinfo  {journal} {Phys.
  Rev. B}\ }\textbf {\bibinfo {volume} {94}},\ \bibinfo {pages} {115416}
  (\bibinfo {year} {2016})}\BibitemShut {NoStop}%
\bibitem [{\citenamefont {Bao}\ and\ \citenamefont {Zhang}(2017)}]{bao17}%
  \BibitemOpen
  \bibfield  {author} {\bibinfo {author} {\bibfnamefont {Z-Q.}\ \bibnamefont
  {Bao}}\ and\ \bibinfo {author} {\bibfnamefont {F.}~\bibnamefont {Zhang}},\
  }\bibfield  {title} {\enquote {\bibinfo {title} {Topological majorana
  two-channel kondo effect},}\ }\href {\doibase 10.1103/PhysRevLett.119.187701}
  {\bibfield  {journal} {\bibinfo  {journal} {Phys. Rev. Lett.}\ }\textbf
  {\bibinfo {volume} {119}},\ \bibinfo {pages} {187701} (\bibinfo {year}
  {2017})}\BibitemShut {NoStop}%
\bibitem [{\citenamefont {van Beek}\ \emph {et~al.}(2018)\citenamefont {van
  Beek}, \citenamefont {Levy~Yeyati},\ and\ \citenamefont
  {Braunecker}}]{vanBeek18}%
  \BibitemOpen
  \bibfield  {author} {\bibinfo {author} {\bibfnamefont {I.~J.}\ \bibnamefont
  {van Beek}}, \bibinfo {author} {\bibfnamefont {A.}~\bibnamefont
  {Levy~Yeyati}}, \ and\ \bibinfo {author} {\bibfnamefont {B.}~\bibnamefont
  {Braunecker}},\ }\bibfield  {title} {\enquote {\bibinfo {title}
  {Nonequilibrium charge dynamics in majorana-josephson devices},}\ }\href
  {\doibase 10.1103/PhysRevB.98.224502} {\bibfield  {journal} {\bibinfo
  {journal} {Phys. Rev. B}\ }\textbf {\bibinfo {volume} {98}},\ \bibinfo
  {pages} {224502} (\bibinfo {year} {2018})}\BibitemShut {NoStop}%
\bibitem [{\citenamefont {Law}\ \emph {et~al.}(2009)\citenamefont {Law},
  \citenamefont {Lee},\ and\ \citenamefont {Ng}}]{law09}%
  \BibitemOpen
  \bibfield  {author} {\bibinfo {author} {\bibfnamefont {K.~T.}\ \bibnamefont
  {Law}}, \bibinfo {author} {\bibfnamefont {P.~A.}\ \bibnamefont {Lee}}, \ and\
  \bibinfo {author} {\bibfnamefont {T.~K.}\ \bibnamefont {Ng}},\ }\bibfield
  {title} {\enquote {\bibinfo {title} {Majorana fermion induced resonant
  andreev reflection},}\ }\href {\doibase 10.1103/PhysRevLett.103.237001}
  {\bibfield  {journal} {\bibinfo  {journal} {Phys. Rev. Lett.}\ }\textbf
  {\bibinfo {volume} {103}},\ \bibinfo {pages} {237001} (\bibinfo {year}
  {2009})}\BibitemShut {NoStop}%
\bibitem [{\citenamefont {Deutscher}\ and\ \citenamefont
  {Feinberg}(2000)}]{deutscher2000}%
  \BibitemOpen
  \bibfield  {author} {\bibinfo {author} {\bibfnamefont {G.}~\bibnamefont
  {Deutscher}}\ and\ \bibinfo {author} {\bibfnamefont {D.}~\bibnamefont
  {Feinberg}},\ }\bibfield  {title} {\enquote {\bibinfo {title} {Coupling
  superconducting-ferromagnetic point contacts by andreev reflections},}\
  }\href {\doibase 10.1063/1.125796} {\bibfield  {journal} {\bibinfo  {journal}
  {Appl. Phys. Lett.}\ }\textbf {\bibinfo {volume} {76}},\ \bibinfo {pages}
  {487--489} (\bibinfo {year} {2000})}\BibitemShut {NoStop}%
\bibitem [{\citenamefont {He}\ \emph {et~al.}(2014)\citenamefont {He},
  \citenamefont {Wu}, \citenamefont {Choy}, \citenamefont {Liu}, \citenamefont
  {Tanaka},\ and\ \citenamefont {Law}}]{he14}%
  \BibitemOpen
  \bibfield  {author} {\bibinfo {author} {\bibfnamefont {J.~J.}\ \bibnamefont
  {He}}, \bibinfo {author} {\bibfnamefont {J.}~\bibnamefont {Wu}}, \bibinfo
  {author} {\bibfnamefont {T.~P.}\ \bibnamefont {Choy}}, \bibinfo {author}
  {\bibfnamefont {X.~J.}\ \bibnamefont {Liu}}, \bibinfo {author} {\bibfnamefont
  {Y.}~\bibnamefont {Tanaka}}, \ and\ \bibinfo {author} {\bibfnamefont {K.~T.}\
  \bibnamefont {Law}},\ }\bibfield  {title} {\enquote {\bibinfo {title}
  {Correlated spin currents generated by resonant-crossed andreev reflections
  in topological superconductors},}\ }\href {\doibase 10.1038/ncomms4232}
  {\bibfield  {journal} {\bibinfo  {journal} {Nat. Commun.}\ }\textbf {\bibinfo
  {volume} {5}},\ \bibinfo {pages} {3232} (\bibinfo {year} {2014})}\BibitemShut
  {NoStop}%
\bibitem [{\citenamefont {Levy~Yeyati}\ \emph {et~al.}(2007)\citenamefont
  {Levy~Yeyati}, \citenamefont {Bergeret}, \citenamefont {Martin-Rodero},\ and\
  \citenamefont {Klapwijk}}]{yeyati07}%
  \BibitemOpen
  \bibfield  {author} {\bibinfo {author} {\bibfnamefont {A.}~\bibnamefont
  {Levy~Yeyati}}, \bibinfo {author} {\bibfnamefont {F.~S.}\ \bibnamefont
  {Bergeret}}, \bibinfo {author} {\bibfnamefont {A.}~\bibnamefont
  {Martin-Rodero}}, \ and\ \bibinfo {author} {\bibfnamefont {T.~M.}\
  \bibnamefont {Klapwijk}},\ }\bibfield  {title} {\enquote {\bibinfo {title}
  {Entangled andreev pairs and collective excitations in nanoscale
  superconductors},}\ }\href {\doibase 10.1038/nphys621} {\bibfield  {journal}
  {\bibinfo  {journal} {Nat. Phys.}\ }\textbf {\bibinfo {volume} {3}},\
  \bibinfo {pages} {455} (\bibinfo {year} {2007})}\BibitemShut {NoStop}%
\bibitem [{\citenamefont {Soori}\ and\ \citenamefont
  {Mukerjee}(2017)}]{soori17}%
  \BibitemOpen
  \bibfield  {author} {\bibinfo {author} {\bibfnamefont {A.}~\bibnamefont
  {Soori}}\ and\ \bibinfo {author} {\bibfnamefont {S.}~\bibnamefont
  {Mukerjee}},\ }\bibfield  {title} {\enquote {\bibinfo {title} {Enhancement of
  crossed andreev reflection in a superconducting ladder connected to normal
  metal leads},}\ }\href {\doibase 10.1103/PhysRevB.95.104517} {\bibfield
  {journal} {\bibinfo  {journal} {Phys. Rev. B}\ }\textbf {\bibinfo {volume}
  {95}},\ \bibinfo {pages} {104517} (\bibinfo {year} {2017})}\BibitemShut
  {NoStop}%
\bibitem [{\citenamefont {Liu}\ \emph {et~al.}(2014)\citenamefont {Liu},
  \citenamefont {Wang},\ and\ \citenamefont {Zhang}}]{liu14}%
  \BibitemOpen
  \bibfield  {author} {\bibinfo {author} {\bibfnamefont {J.}~\bibnamefont
  {Liu}}, \bibinfo {author} {\bibfnamefont {J.}~\bibnamefont {Wang}}, \ and\
  \bibinfo {author} {\bibfnamefont {F.-C.}\ \bibnamefont {Zhang}},\ }\bibfield
  {title} {\enquote {\bibinfo {title} {Controllable nonlocal transport of
  majorana fermions with the aid of two quantum dots},}\ }\href {\doibase
  10.1103/PhysRevB.90.035307} {\bibfield  {journal} {\bibinfo  {journal} {Phys.
  Rev. B}\ }\textbf {\bibinfo {volume} {90}},\ \bibinfo {pages} {035307}
  (\bibinfo {year} {2014})}\BibitemShut {NoStop}%
\bibitem [{\citenamefont {Chen}\ \emph {et~al.}(2015)\citenamefont {Chen},
  \citenamefont {Shi},\ and\ \citenamefont {Xing}}]{Chen2015}%
  \BibitemOpen
  \bibfield  {author} {\bibinfo {author} {\bibfnamefont {W.}~\bibnamefont
  {Chen}}, \bibinfo {author} {\bibfnamefont {D.~N.}\ \bibnamefont {Shi}}, \
  and\ \bibinfo {author} {\bibfnamefont {D.~Y.}\ \bibnamefont {Xing}},\
  }\bibfield  {title} {\enquote {\bibinfo {title} {Long-range cooper pair
  splitter with high entanglement production rate},}\ }\href {\doibase
  10.1038/srep07607} {\bibfield  {journal} {\bibinfo  {journal} {Sci. Rep.}\
  }\textbf {\bibinfo {volume} {5}},\ \bibinfo {pages} {7607} (\bibinfo {year}
  {2015})}\BibitemShut {NoStop}%
\bibitem [{\citenamefont {Rokhinson}\ \emph {et~al.}(2012)\citenamefont
  {Rokhinson}, \citenamefont {Liu},\ and\ \citenamefont {Furdyna}}]{rokhi2012}%
  \BibitemOpen
  \bibfield  {author} {\bibinfo {author} {\bibfnamefont {L.~P.}\ \bibnamefont
  {Rokhinson}}, \bibinfo {author} {\bibfnamefont {X.}~\bibnamefont {Liu}}, \
  and\ \bibinfo {author} {\bibfnamefont {J.~K.}\ \bibnamefont {Furdyna}},\
  }\bibfield  {title} {\enquote {\bibinfo {title} {The fractional a.c.
  josephson effect in a semiconductor-superconductor nanowire as a signature of
  majorana particles},}\ }\href {\doibase 10.1038/nphys2429} {\bibfield
  {journal} {\bibinfo  {journal} {Nat. Phys.}\ }\textbf {\bibinfo {volume}
  {8}},\ \bibinfo {pages} {795} (\bibinfo {year} {2012})}\BibitemShut {NoStop}%
\bibitem [{\citenamefont {Kwon}\ \emph {et~al.}(2004)\citenamefont {Kwon},
  \citenamefont {Sengupta},\ and\ \citenamefont {Yakovenko}}]{kwon04}%
  \BibitemOpen
  \bibfield  {author} {\bibinfo {author} {\bibfnamefont {H.-J.}\ \bibnamefont
  {Kwon}}, \bibinfo {author} {\bibfnamefont {K.}~\bibnamefont {Sengupta}}, \
  and\ \bibinfo {author} {\bibfnamefont {V.~M.}\ \bibnamefont {Yakovenko}},\
  }\bibfield  {title} {\enquote {\bibinfo {title} {Fractional ac josephson
  effect in p- and d-wave superconductors},}\ }\href {\doibase
  10.1140/epjb/e2004-00066-4} {\bibfield  {journal} {\bibinfo  {journal} {Eur.
  Phys. J. B}\ }\textbf {\bibinfo {volume} {37}},\ \bibinfo {pages} {349--361}
  (\bibinfo {year} {2004})}\BibitemShut {NoStop}%
\bibitem [{\citenamefont {Cayao}\ \emph {et~al.}(2017)\citenamefont {Cayao},
  \citenamefont {San-Jose}, \citenamefont {Black-Schaffer}, \citenamefont
  {Aguado},\ and\ \citenamefont {Prada}}]{cayao17}%
  \BibitemOpen
  \bibfield  {author} {\bibinfo {author} {\bibfnamefont {J.}~\bibnamefont
  {Cayao}}, \bibinfo {author} {\bibfnamefont {P.}~\bibnamefont {San-Jose}},
  \bibinfo {author} {\bibfnamefont {A.~M.}\ \bibnamefont {Black-Schaffer}},
  \bibinfo {author} {\bibfnamefont {R.}~\bibnamefont {Aguado}}, \ and\ \bibinfo
  {author} {\bibfnamefont {E.}~\bibnamefont {Prada}},\ }\bibfield  {title}
  {\enquote {\bibinfo {title} {Majorana splitting from critical currents in
  josephson junctions},}\ }\href {\doibase 10.1103/PhysRevB.96.205425}
  {\bibfield  {journal} {\bibinfo  {journal} {Phys. Rev. B}\ }\textbf {\bibinfo
  {volume} {96}},\ \bibinfo {pages} {205425} (\bibinfo {year}
  {2017})}\BibitemShut {NoStop}%
\end{thebibliography}%

\end{document}